# Universal Properties of Cuprate Superconductors: $T_c$ Phase Diagram, Room-Temperature Thermopower, Neutron Spin Resonance, and STM Incommensurability Explained in Terms of Chiral Plaquette Pairing

Jamil Tahir-Kheli* and William A. Goddard III*

*Materials and Process Simulation Center (MC 139-74)*

*California Institute of Technology, Pasadena CA 91125*

*To whom correspondence should be addressed. Fax: 626-585-0918. Tel: 626-395-8148. Email: jamil@wag.caltech.edu, wag@wag.caltech.edu

ABSTRACT We report that four properties of cuprates and their evolution with doping are consequences of simply counting four-site plaquettes arising from doping: (1) the universal $T_c$ phase diagram (superconductivity between ≈0.05 and ≈0.27 doping per $CuO_2$ plane, and optimal $T_c$ at ≈0.16), (2) the universal doping dependence of the room-temperature thermopower, (3) the superconducting neutron spin resonance peak (the "41 meV peak"), and (4) the dispersionless scanning tunneling conductance incommensurability. Properties (1), (3), and (4) are explained with no adjustable parameters, and (2) is explained with exactly one. The successful quantitative interpretation of four very distinct aspects of cuprate phenomenology by a simple counting rule provides strong evidence for four-site plaquette percolation in these materials. This suggests that inhomogeneity, percolation, and plaquettes play an essential role in cuprates. This geometric analysis may provide a useful guide to search for new compositions and structures with improved superconducting properties.

We provide a simple explanation of the universal dependence of four properties of cuprate superconductors on doping (x) in the $CuO_2$ planes: (1) the universal dependence of $T_c$ (superconductivity between x≈0.05 and ≈0.27 doping per $CuO_2$ plane, and optimal $T_c$ at ≈0.16),[1,2] (2) the room-temperature thermopower (Seebeck effect),[1-5] (3) the neutron spin ($\pi,\pi,\pi$) resonance peak,[6-22] and (4) the non-dispersing conductance incommensurabilities in STM (observed thus far only for single-layer Bi2201).[23,24]

It is hard to imagine four experiments that are more different. The $T_c$ phase diagram is due to the nature of the superconducting pairing and its doping evolution, the universal thermopower is observed in the normal state near room-temperature and relates simultaneous heat and charge transport, the neutron resonance probes spin fluctuations, and the STM measures local density of states (LDOS) variations on an atomic scale. We explain all four experiments here using simple counting arguments.

It is well known that the superconducting critical temperature, $T_c$, for all cuprates fits the expression, $(T_c/T_{c,max}) \approx 1 - 82.6(x-0.16)^2$, where x is the hole doping per Cu in the $CuO_2$ planes.[1,2] This leads to the three universal doping values where superconductivity first appears at x≈0.05, is optimal at x≈0.16, and disappears above x≈0.27. This remarkable universality has not been explained.

The room-temperature (290K) thermopower, or Seebeck effect, for all cuprates decreases strongly with increased doping (from +80 to –13 $\mu$V/K) with the same universal dependence. In addition, the temperature dependence at high temperatures is anomalous. Rather than S=BT, as expected from entropy transport due to electrons in metals (the Mott formula), all cuprates have the form $A+BT$ where $A$ is large and strongly doping dependent while $B$ is doping independent.[1-5]

The neutron spin inelastic scattering shows a very strong resonance near "41 meV" at the AF wavevector ($\pi,\pi$) [($\pi,\pi,\pi$) in bi-layer materials] that is nearly the same for all cuprates, but doping dependent while tracking $T_c$.[25,26] The peak occurs at the center of an "hourglass" dispersion with the high energy sheet doping independent and the lower sheet doping and material dependent.[27-30]

The recently observed doping dependent STM incommensurability in single-layer Bi2201[23,24] is anomalous because the wavelength increases with increasing doping rather than decreasing with increasing doping as expected from the mean separation of holes.

No theory has yet explained all four within a single framework. Electronic and small polaron models have been proposed for the $T_c$ phase diagram[31-33] and spin-vortex and stripe models for the neutron resonance.[21,34,35] We show here, that simple counting combined with a few simple assumptions regarding the character of the doping, *quantitatively* explains the thermopower with exactly one adjustable parameter while simultaneously explaining the other three experiments with no adjustable parameters. Our assumptions are derived from the results of ab-initio B3LYP DFT calculations on undoped and explicitly doped $La_{2-x}Sr_xCuO_4$[36,37] and our Chiral Plaquette Pairing model (CPP).[38,39]

Our model for doping rests on three assumptions. First, doping leads to a hole in an out-of-the-plane impurity orbital with $Op_z$-$Cud_{z^2}$-$Op_z$ character that is orthogonal to the planar Cu/O $x^2-y^2/p\sigma$ band. Our QM calculations[36-39] show that this orbital is delocalized over a four-site square Cu plaquette in the vicinity of the dopant and is comprised (for LSCO) predominantly of apical O $p_z$ (above Cu in $CuO_2$ plane) and Cu $d_{z^2}$ hole character. We refer to the four Cu sites included in the four-site plaquettes as doped sites. The undoped Cu sites remain localized $d^9$ states with neighboring AF coupling $J_{dd}$=0.13 eV = 130 meV (the value found in undoped materials[40,41]). Figure 1 shows a 2D snapshot at doping x=0.16.

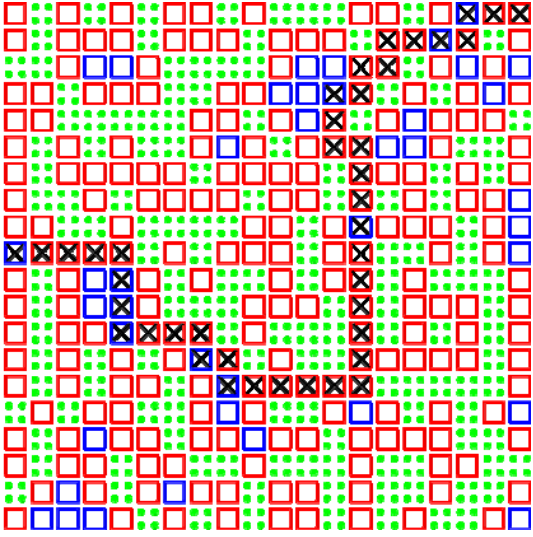

**Figure 1.** (a) Illustration of the three types of electrons in a single $CuO_2$ plane for x=0.16 doping. Green dots indicate undoped Cu $d^9$ spins that are AF coupled. Each square indicates a four-site plaquette centered at a dopant and comprised of out-of-plane orbitals (apical O $p_z$ and Cu $dz^2$). The dopants are distributed randomly within the condition of even separations. The red (blue) squares are surface (interior) plaquettes. A Cu/O $x^2$-$y^2$/$p\sigma$ metallic band forms inside the percolating region of the plaquettes due to Cu $x^2$-$y^2$ orbital energy lowering relative to O $p\sigma$ as shown in figure 2. The crosses schematically show one percolating path. Optimal doping, x=0.16, is shown.

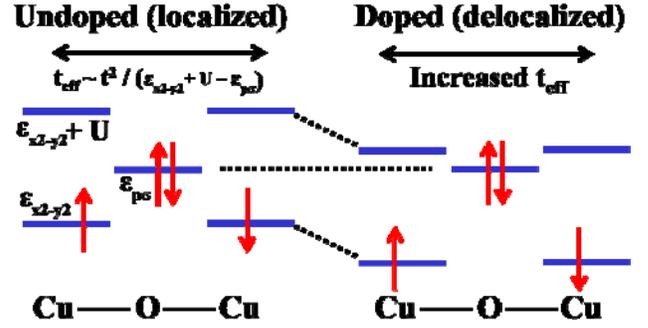

**Figure 2.** Localized hole formation in out-of-the-plane Cu $dz^2$ and apical O $p_z$ orbitals leads to delocalization of the Cu $d^9$ spins within the plaquette swath due to increased Cu-Cu hopping in the $CuO_2$ plane arising from the stabilization of the Cu $dx^2$-$y^2$ orbital energy relative to the planar O $p\sigma$. In the undoped regions, localized Cu $d^9$ spins remain.

We assume that placing dopants on neighboring sites leads to repulsive interaction so that there are no plaquette overlaps. For the calculations in this paper, we assume a more restrictive doping for which the plaquette centers are always separated by an even number of Cu-Cu lattice spacings as shown in figure 1. This allows doping up to x=0.25 with no plaquette overlap and leads to analytic expressions for the quantities of interest. This restriction does not alter the results in this paper (see Supporting Information).

Second, as doping increases, there is a critical concentration above which the four-site plaquettes percolate through the crystal. In this regime we expect that the planar Cu/O $x^2$-$y^2$/$p\sigma$ orbitals on the doped sites inside the percolating swath delocalize to form the standard Cu/O $x^2$-$y^2$/$p\sigma$ metallic band. This is because the Cu $x^2$-$y^2$ orbital energy is lowered relative to the O $p\sigma$ due to the reduced Coulomb repulsion from the hole in the out-of-the-plane plaquette state. This increases the effective Cu-Cu hopping relative to the undoped sites and leads to delocalization as shown in figure 2.

These two assumptions lead to three types of electrons: (1) the undoped Cu $d^9$ spins (dots in figure 1) that are coupled to each other with the undoped AF $J_{dd}$=130 meV, (2) the Cu/O $x^2$-$y^2$/$p\sigma$ metallic band inside the percolating plaquette region (inside squares in figure 1), and (3) the hole comprising the four-site plaquette orbital that is orthogonal to the Cu/O $x^2$-$y^2$/$p\sigma$ metallic band (shown as squares in figure1). Figure 1 also distinguishes surface plaquettes in contact with undoped $d^9$ spins (red squares) and interior plaquettes not in contact with $d^9$ spins (blue squares).

Third, we assume superconducting pairing occurs only for surface plaquettes (adjacent to undoped $d^9$ sites). We argued previously[38,39] that this occurs because the interaction with the $d^9$ spins makes these plaquettes chiral, but such detail is not necessary for the results presented here.

On the basis of the above assumptions, the onset of superconductivity occurs when the four-site plaquettes percolate in three-dimensions (3D) because a metallic band is formed on the percolating swath. Using a linear algorithm[42], we calculate that 3D percolation occurs at x=0.066 holes per $CuO_2$ plane (experimental value 0.053[43]). Similar calculations assuming a staggered $CuO_2$ planar structure in $La_{2-x}Sr_xCuO_4$ lead to ≈0.05.

It has recently been shown in LSCO[43] that the $T_c$ at the transition to superconductivity does not start at $T_c$=0, but instead jumps to a finite value of ≈3.5K. In our model, the number of superconducting electrons is 4x and the strength of the pairing is proportional to the number of surface plaquettes. Thus at the point at which percolation is achieved, x=0.053, the strength of the pairing should be proportional to 4x = 0.2, leading to a jump to finite $T_c$ at the transition consistent with experiment.

At doping x=0.25, our model implies that no undoped Cu $d^9$ sites are left to induce superconducting pairing. This leads to x=0.25 as the maximum doping for superconductivity (experimental value ≈0.27). See Supporting Information for further details.

The optimal $T_c$ occurs when the pairing maximizes the energy lowering. This occurs when the number of electrons that can pair [the density of states at the Fermi level, N(0)] times the pairing energy, V, is maximized. Thus optimal $T_c$ occurs when N(0)V is maximized.

In our model, N(0) ~ $\Omega_M/\Omega_{total}$, where $\Omega_M$ is the number of metallic sites (doped Cu sites) and $\Omega_{total}$ is the total sites. $\Omega_M/\Omega_{total}$=4x since each plaquette adds four Cu sites to the metallic swath. The pairing strength is the ratio of the number of surface plaquettes, $S_p$, to the number of metallic sites, V ~ $S_p/\Omega_M$, leading to N(0)V ~ $S_p/\Omega_{total}$.

With random doping, the probability that a plaquette is surrounded by four plaquettes is $(4x)^4$. The probability the plaquette is on the surface (has at least one AF $d^9$ neighbor) is 1−$(4x)^4$ leading to $S_p/\Omega_{total}$=x[1−$(4x)^4$]. The maximum occurs at x=1280$^{-1/4}$≈0.167 (experimental value ≈0.16). The optimal doping value does not depend on differences of the Cooper pairing temperature, $T_p$, and the pair phase coherence temperature, $T_\varphi$, because generally $T_c$=min($T_p$, $T_\varphi$), and optimal doping occurs at the crossover from the phase fluctuation to pairing regime, $T_p$=$T_\varphi$[44]. This explains the three universal doping values of the superconducting phase.

The undoped $d^9$ clusters in figure 1 (green dots) are described by the Heisenberg AF spin Hamiltonian with the undoped AF coupling, $J_{dd}$=130 meV. These finite AF clusters will spin couple to neighboring AF clusters through the metallic $x^2$-$y^2$/$p\sigma$ electrons and by coupling with the surface plaquette hole spins. The random locations of the surface plaquettes lead to a disordered magnet with a finite spin correlation length, $\xi$. The $(\pi,\pi)$ state [or $(\pi,\pi,\pi)$ state in bi-layer systems] has zero excitation energy in the infinite 2D Heisenberg antiferromagnet, but will be gapped due to the disorder. We estimate the excitation energy, $E_{res}$, by $E_{res} = \hbar c_{sw}/2\xi$, where $c_{sw}$ is the undoped AF magnon spin velocity. This is derived by applying the uncertainty principle, $\Delta x \Delta p = \hbar/2$, to the spin-wave dispersion, $\hbar\omega = c_{sw}\Delta p$, and choosing $\Delta p = \hbar/2\xi$.

**Table I.** Calculated neutron resonance peak energy versus experiment for under, optimal, and overdoped (UD, OP, OD) Bi2212, YBCO, and Tl2201.[a]

| Material | $T_c$ (K) | Doping (x) | $S_p/\Omega_{tot}$ | Res Energy (meV) | |
|---|---|---|---|---|---|
| | | | | Expt | Theory |
| YBCO$_{6.5}$ (UD) | 52 | 0.087 | 0.086 | 25 | 31.7 |
| YBCO$_{6.7}$ (UD) | 67 | 0.102 | 0.099 | 33 | 34.1 |
| YBCO$_{6.9}$ (OP) | 93 | 0.167 | 0.133 | 41 | 39.6 |
| Tl2201 (OP) | 90 | 0.167 | 0.133 | 47 | 39.6 |
| Bi2212 (OP) | 91 | 0.167 | 0.133 | 43 | 39.6 |
| Bi2212 (OD) | 87 | 0.183 | 0.130 | 42 | 39.1 |
| Bi2212 (OD) | 83 | 0.193 | 0.124 | 38 | 38.2 |
| Bi2212 (OD) | 70 | 0.213 | 0.101 | 34 | 34.4 |

[a]The data is from references[9,10,15]. We used $T_{c,max}$ = 93, 90, and 91K for YBCO, Tl2201, and Bi2212 respectively to obtain the doping, x, in column 3 by applying $(T_c/T_{c,max}) \approx 1-82.6(x-0.16)^2$. The number of surface plaquettes is given by $S_p/\Omega_{tot} = x[1-(4x)^4]$. $E_{res} = \hbar c_{sw}/2\xi$ where $\xi = (S_p/\Omega_{tot})^{-1/2}a$ and $\hbar c_{sw} = 2^{1/2} Z_c J_{dd} a$ with $Z_c \approx 1.18$[45].

The spin-wave velocity, $c_{sw}$, is determined from the undoped AF coupling, $J_{dd}$, and harmonic spin-wave expansions[45], $\hbar c_{sw} = \sqrt{2} Z_c J_{dd} a$, where a is the Cu–Cu lattice distance and $Z_c \approx 1.18$. The correlation length, $\xi$, is the mean spacing between surface plaquettes, $\xi = (S_p/\Omega_{tot})^{-1/2}a$, and is known as a function of doping, x, as shown above, while x can be obtained from the universal doping $T_c/T_{c,max}$ equation[1,2] as described above. Thus the neutron resonance energy, $E_{res}$, is completely determined with no adjustable parameters. Since $\xi$ is shortest at optimal $T_c$, the resonance peak tracks $T_c$ rather than increasing for increasing doping away from the undoped AF phase.[25,26]

Table 1 compares our calculated resonance peak energy with experiment for underdoped, optimally doped, and overdoped cuprates YBCO, Bi-2212, and Tl-2201. The fit is very good.

The energy integrated neutron peak susceptibility, $\int d\omega \chi''(q,\omega)$ where $q=(\pi,\pi,\pi)$, is known experimentally and can be estimated by summing the contribution to the integral from each finite cluster. This estimate is reasonable since the correlation length is on the order of the cluster sizes.

We computed the S=0 ground state and S=1 first excited state for all AF cluster shapes and sizes up to 24 spins (24 AF spins required Lanczos diagonalization over $\approx 2.7 \times 10^6$ states) along with the corresponding spin-flip matrix elements to obtain an energy integrated spectral weight of $5.1\mu_B^2$ per f.u. for optimal doping (see Supporting Information). Experiment finds 1.9 for optimally doped Bi-2212 and 1.6 YBCO[9,11]. Our estimated result is approximately 2.5 times larger than experiment. Generally, models that attribute the resonance to Fermi surface effects are an order of magnitude smaller. A phenomenological $(\pi,\pi)$ spin-fluctuation enhancement is invoked to scale the value up to experiment.

**Table II.** Comparison of calculated conductance incommensurability versus data on Bi2201.[a]

| $T_c$(K) | Doping (x) | Incommensurability ($\lambda/a$) | |
|---|---|---|---|
| | | Expt[23,24] | Theory |
| 25 | 0.100 | 4.5 ± 0.2 | 4.3 |
| 32 | 0.128 | 5.1 ± 0.2 | 5.1 |
| 35 | 0.160 | 6.2 ± 0.2 | 6.6 |

[a]$\lambda$ is the incommensurability and a is the planar Cu–Cu separation. The theoretical expression is $2/(1-4x)+1$.

The experimental neutron resonance width, $\Gamma$, is found to be resolution limited (~5 meV) in YBCO and slightly broader than resolution for Bi2212 in the superconducting state[9,11,15]. This peak is substantially broadened in the normal state. We have calculated $\Gamma$ at 20K and at $T_c$ for optimally doped YBCO and Bi-2212 using angle-resolved photoemission (ARPES) band structures for Bi-2212[15,46] with an STM gap of 41.5 eV and from tight-binding models for YBCO[47-49] with an STM gap of 20.0 eV. For YBCO, we calculate $\Gamma$=1.3 and 45.3 meV at 20K and 92K, and for Bi-2212, $\Gamma$=0.1 and 22.8 meV at 20K and 92K (see Supporting Information for details). Our calculated peaks are resolution limited in the superconducting phase and are substantially broadened in the normal state (factor of $\approx$30). The peak width for Bi-2212 is observed to be slightly broader than instrument resolution[15]. This may be due to spatial variation of the gap.

The wavelength of the STM is given by the expected size of the metallic region between $d^9$ sites. The expected number of doped plaquettes between two $d^9$ regions is $1/(1-4x)$ since $1-4x$ is the probability that a plaquette is undoped ($d^9$). This leads to a total wavelength of $[2/(1-4x)]+1$ in units of the Cu–Cu lattice spacing, a, since each plaquette is 2x2 Cu sites and one further step is needed to get back to a $d^9$ site. Since the incommensurability is structural, it should be independent of the voltage. Table 2 shows the good fit to experiment.[23,24]

An alternative explanation by Wise et al[23] is that this incommensurability arises from a charge density wave due to the Fermi surface nesting vector near $(\pi,0)$ that decreases as hole doping increases.

Figure 3 shows that the universal room-temperature thermopower, S(290K), decreases as a function of doping.[1,2] The electronic thermopower, $S_e$, due to the $x^2-y^2/p\sigma$ metallic electrons, leads to a linear temperature dependence with the magnitude depending on the derivative of the logarithm of the density states (DOS) and scattering time, $\tau$ (Mott formula[50,51]). Any change in the DOS or $\tau$ by a constant factor due to doping does not change $S_e$. Thus $S_e$ at 290K is doping independent and cannot account for the observed doping dependent room-temperature thermopower. Experiments in the non-superconducting region, x>0.27, find $S_e \approx -13$ µV/K.

Besides $S_e$, the plaquette model can lead to an additional contribution to the thermopower from the magnon drag arising from the non-equilibrium distribution of heat-carrying magnons in the undoped $d^9$ AF regions. By analogy to phonon drag,[50-52] we expect that $S_{mag} = f(mc^2/e)(\tau_{mag}/\tau_e)(1/T)$, where m is the electron mass, e is the charge, c is the magnon spin wave velocity, f is the fraction of $x^2-y^2/p\sigma$ band momentum dissipated into magnons, $\tau_{mag}$ is the magnon scattering time, $\tau_e$ is the band scattering time, and T is temperature. At room-temperature, $1/\tau_e \sim T$ and $f \sim S/\Omega_M$ is the ratio of the surface area to the metallic swath. The magnons dissipate their momentum primarily by impurity scattering with the surface plaquettes and the metallic electrons. Thus $1/\tau_{mag} \sim S/\Omega_{AF}$ is the ratio of the surface area of the $d^9$ regions to its size. This leads to $S_{mag} \sim \Omega_{AF}/\Omega_M = (1-4x)/4x$. The surface term and T cancels and leads to a constant.

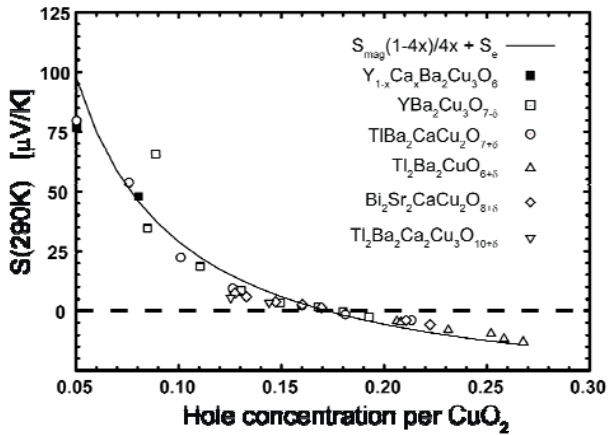

**Figure 3.** Fitting the thermopower at 290K as a function of hole doping. The experimental data is from references.[1,2] The solid line is the prediction from the plaquette theory, $S(290K) = S_{mag}(1-4x)/4x + S_e$ where $S_{mag}=27.6\mu V/K$ and $S_e=-13\mu V/K$. $S_{mag}$ is the only adjustable parameter.

Combining these terms, we get $S(290K)=S_{mag}(1-4x)/4x+S_e$, where $S_{mag}$ is an undetermined constant (best fit is 27.6μV/K) and $S_e \approx -13\mu V/K$ from experiment. Figure 3 shows the good fit of this expression for $x>0.05$ where the metallic phase begins due to plaquette percolation.

In conclusion, we show that the simple assumption that dopants in cuprates lead to the formation of four-site plaquettes localized in the vicinity of the dopants with a Cu/O $x^2-y^2/p\sigma$ metallic band created in the percolating region explains quantitatively the doping evolution of four universalities of cuprates. They are: (1) the critical values for superconductivity (onset, optimal, and maximum doping), (2) the "41 meV" neutron spin resonance peak, due to the finite correlation length of the AF regions between the percolating plaquette swaths, (3) the dispersionless STM incommensurability, and (4) the room-temperature thermopower due to magnon drag in the AF regions. The first three results are obtained with no adjustable parameters. The fourth is explained with exactly one adjustable parameter to fit experiment.

We believe the success in explaining four important, but seemingly unrelated properties of the cuprates using only counting arguments is strong evidence for the role of four-site lattice plaquette percolation in cuprates. Since increasing the surface area to volume ratio increases $T_c$, we suggest higher $T_c$ may be obtained by controlling the location of dopants.

ACKNOWLEDGEMENT The authors acknowledge discussions with V. Hinkov. Support for this research was provided by DOD-DARPA (0211720) and ONR-PROM (N00014-06-1-0938). The computational facilities at the Materials and Process Simulation Center were provided by ARO-DURIP and ONR-DURIP.

SUPPORTING INFORMATION AVAILABLE Discussion of the general doping of plaquettes. Computational details of energy integrated neutron susceptibility and spin resonance width. This information is available free of charge via the Internet at http://pubs.acs.org.

# Universal Properties of Cuprate Superconductors: T$_c$ Phase Diagram, Room-Temperature Thermopower, Neutron Spin Resonance, and STM Incommensurability Explained in Terms of Chiral Plaquette Pairing

Jamil Tahir-Kheli and William A. Goddard III

*Materials and Process Simulation Center (MC 139-74)*
*California Institute of Technology, Pasadena CA 91125*

jamil@wag.caltech.edu, wag@wag.caltech.edu

## Supporting Information

**Doping restrictions and its effect on the optimal doping value**

Figure S1(a) shows the restricted plaquette doping used in this paper and figure S1(b) shows the more general doping.

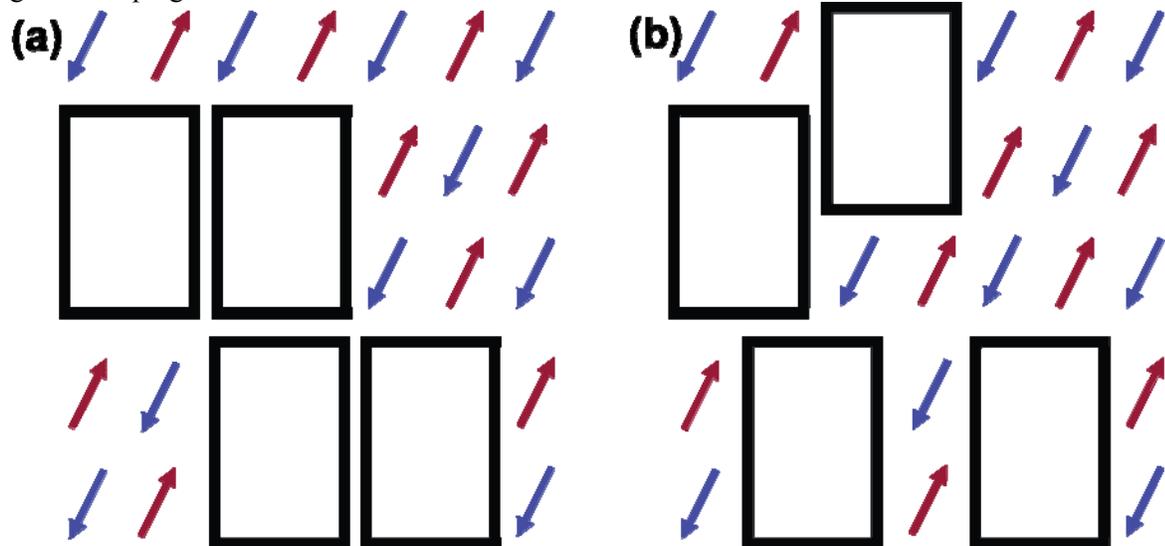

**Figure S1.** (a) restricted plaquette doping where the plaquettes centers are separated by an even number of lattice spacings. (b) general plaquette doping with no overlap. This constraint can be fulfilled up to 0.187 doping.

The restricted doping in figure S1(a) can be doped all the way to x=0.25 where there are no remaining $d^9$ sites. For random doping of plaquettes in the general form of figure S1(b), we find that for x>0.187, it is not possible to satisfy the constraint of no overlap of the plaquettes. We calculate this value by randomly doping a large ensemble of 1000 X 1000 lattices and computing the doping where the constraint cannot be satisfied. This doping is significantly greater than our calculated optimal doping of 0.167.

The detailed location of the dopants in the cuprates is determined by details of the high-temperature fabrication and the annealing/quenching profile. For dopings less than 0.187, the Coulomb repulsion of the impurity dopants will lead to patches of restricted plaquettes as shown in figure S1(a) that are joined to patches shifted by one lattice spacing. Since the calculated optimal doping at 0.167 is well within the regime where four-site plaquettes can be formed without overlap, our estimate of the optimal doping value is reliable.

It is interesting to continue doping beyond the plaquette overlap constraint for x<0.187. Assuming that the Coulomb repulsion of the dopant impurities will lead to further four-site plaquettes at locations that cover three $d^9$ spins, this constraint can be satisfied up to 0.226. Above x=0.226, the impurities lead to additional plaquettes that cover two $d^9$ spins. This constraint is exhausted at x=0.271. Finally, one-spin plaquettes can be doped up to x=0.317.

In the paper, we argued that at x=0.25 there are no longer any remaining $d^9$ spins to lead to superconducting pairing. In the more general doping picture described here, pairing will disappear when only isolated $d^9$ spins remain because the nearest-neighbor AF correlation ($J_{dd}$) that leads to pairing is lost. This occurs at x=0.271. Experiment is ≈0.27.

For the neutron resonance, thermopower, and STM incommensurability, the surface area of the plaquettes is unnecessary. Only average values for the spacing between plaquettes and $d^9$ regions are used. Thus the main results of the paper are not dependent on the details of the plaquette doping.

**Integrated neutron spin susceptibility**

The imaginary part of the magnetic susceptibility due to AF clusters of size n is given by $\chi''(q,\omega)$ where[1]

$$\frac{\chi_n''(q,\omega)}{(g\mu_B)^2} = \frac{\pi}{2} \cdot \frac{N_{clus}(n)}{N_{cells}} |<1|S^+(q)|0>|^2 \, \delta(\hbar\omega - E_n).$$

$N_{clus}(n)$ is the number of clusters of size n, $N_{cells}$ is the total cells, $E_n$ is the cluster energy gap, g=2 is the g-factor, and $S^+(q) = \sum e^{iqR} S^+(R)$ is the Fourier transformed spin raising operator summed over the sites R in the finite cluster. The matrix element is between the ground state (S=0) and the lowest excited state (S=1). The units are $\mu_B^2$/eV/f.u. Integrating over energy removes the delta function in the above expression. The energy integrated susceptibility is the sum over all clusters

$$\int d\omega \chi''(q,\omega) = \sum_n \int d\omega \chi_n''(q,\omega).$$

To determine the matrix element in the equation, we computed the S=0 ground state and S=1 first excited state for all AF cluster shapes and sizes up to 24 spins. The four-site plaquette doping as shown in figure S1(a) leads to finite clusters comprised of 4, 8, 12, 16, … $d^9$ sites. The nearest-neighbor AF spin coupling is the same as for undoped systems, $J_{dd}$ = 130 meV. There are no periodic boundary conditions applied to calculate the ground state to first excited state energy splitting.

We computed the energy splitting for all possible AF clusters up to 24 spin sites. The ground state is determined by diagonalizing the S=0 spin state that has 2,704,156 terms. Lanczos diagonalization is used. This leads to the ground state energy. Since the neutron resonance is a spin-flip scattering, the lowest energy state with S=1 is the excited state that is probed by the neutrons. A second Lanczos diagonalization to determine the ground state in the state space of S=1 with 2,496,144 states is performed. The difference is the energy splitting.

Figure S2 shows all the possible AF $d^9$ clusters possible up to 20 sites and figure S3 shows all the 24-site clusters. For each size, the probability of each cluster shape was determined in addition to the S=0 ground state to S=1 energy splitting. The probabilities and energies were used to determine the mean energy splitting for each cluster size.

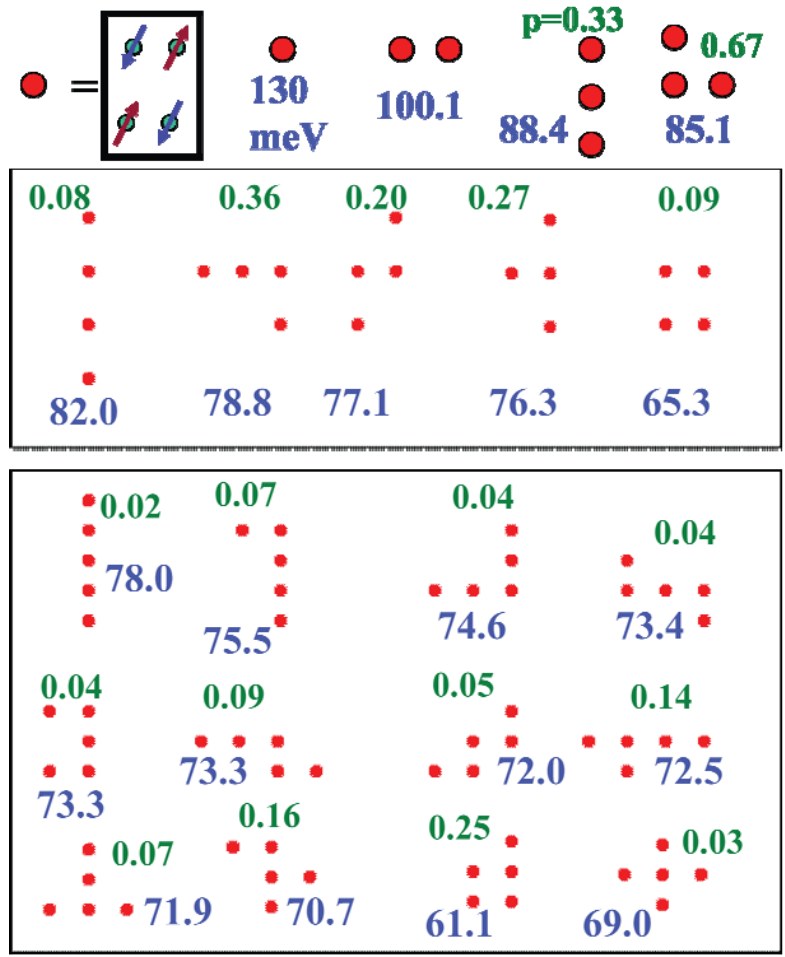

**Figure S2.** All 4, 8, 12, 16, and 20 AF d⁹ site clusters. The red dot expands to four d⁹ spins as shown in the upper left corner. The ground state to S=1 excited state energy is shown in meV in blue and the probability of the cluster is shown in green. The undoped spin-spin coupling, $J_{dd}$ = 130 meV, is used in these calculations. There is only one 4-site and 8-site cluster. There are 2, 5, and 12 clusters for 12, 16, and 20 sites, respectively.

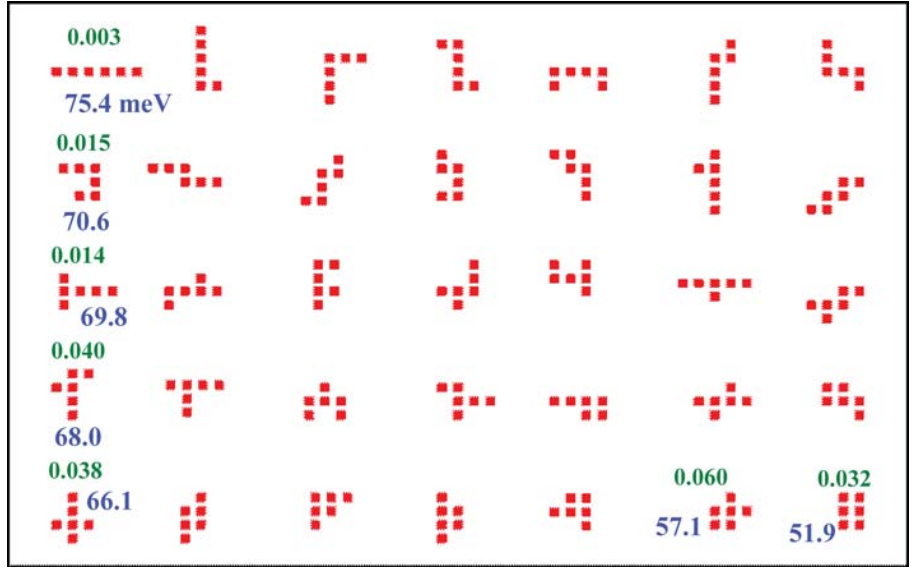

**Figure S3.** The 35 possible 24-site AF d$^9$ spin clusters. The energy splitting is shown in blue and the probability is in green. The most probable cluster is the second to last cluster on the lower right with energy splitting of 57.1 meV and probability 0.06. The clusters are arranged from highest energy splitting to lowest.

24-site AF d$^9$ clusters are too small to include all the cluster sizes that can appear for the superconducting range of dopings. Thus we must extrapolate our energies to larger clusters.

Since the spin-wave dispersion is linear in momentum, $\omega \sim k$, and the smallest k in a finite cluster is $\sim N^{-1/2}$, where N is the cluster size, we expect our calculated energy splittings to be fitted by a power series in $N^{-1/2}$. The mean energy as a function of N and a fit to the data with the first two terms in the power series in $N^{-1/2}$ is shown in figure S4. The fit is excellent with an RMS error of 0.61 meV and a maximum percentage error of 1.03%.

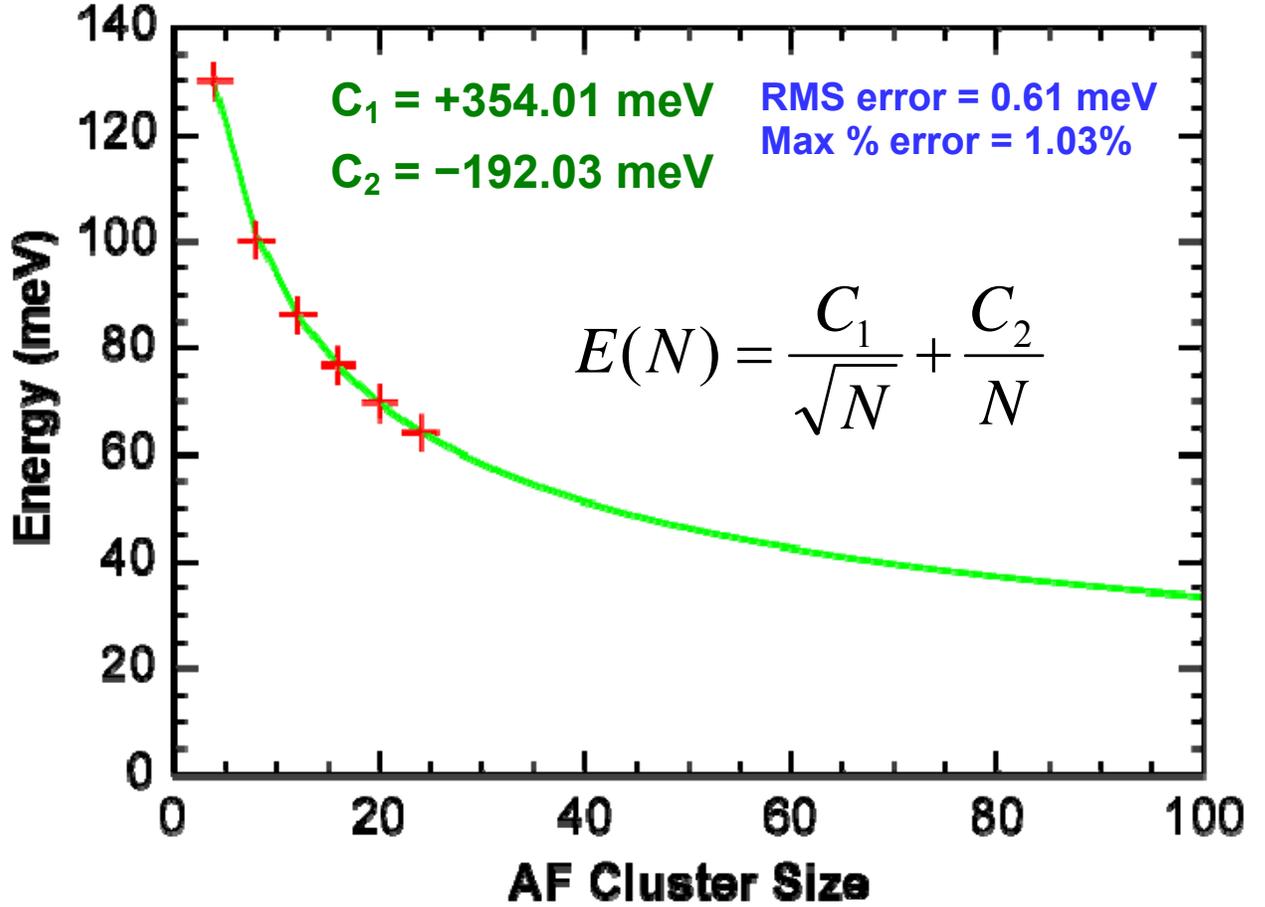

**Figure S4.** Fit of the mean energy for cluster sizes up to 24-sites with the first two terms in a power series expansion in $N^{-1/2}$ where N is the cluster size. An $N^{-1/2}$ series expansion is expected from spin-wave theory. The fit is excellent using only the first two terms in the series. The maximum error is 1.03 meV and the RMS error is 0.61 meV. The largest percentage error of the six data points is 1.03%. Note that the coefficient $C_2$ is negative.

From spin-wave theory, the matrix element, $\frac{1}{n}|<1|S^+(q)|0>|^2$, increases to infinity with increasing n. We used our eigenstates for clusters up to 24 AF spins to calculate this term for $q=(\pi,\pi)$. This term fits the power law $Cn^\alpha$ with an RMS error of 0.01 and maximum percentage error of 0.59%, where C=0.86 and α=0.32 as shown in figure S5. An interesting observation is that the expression $CN^\alpha$ is very close to $(4/3)(N/4)^{1/3}$.

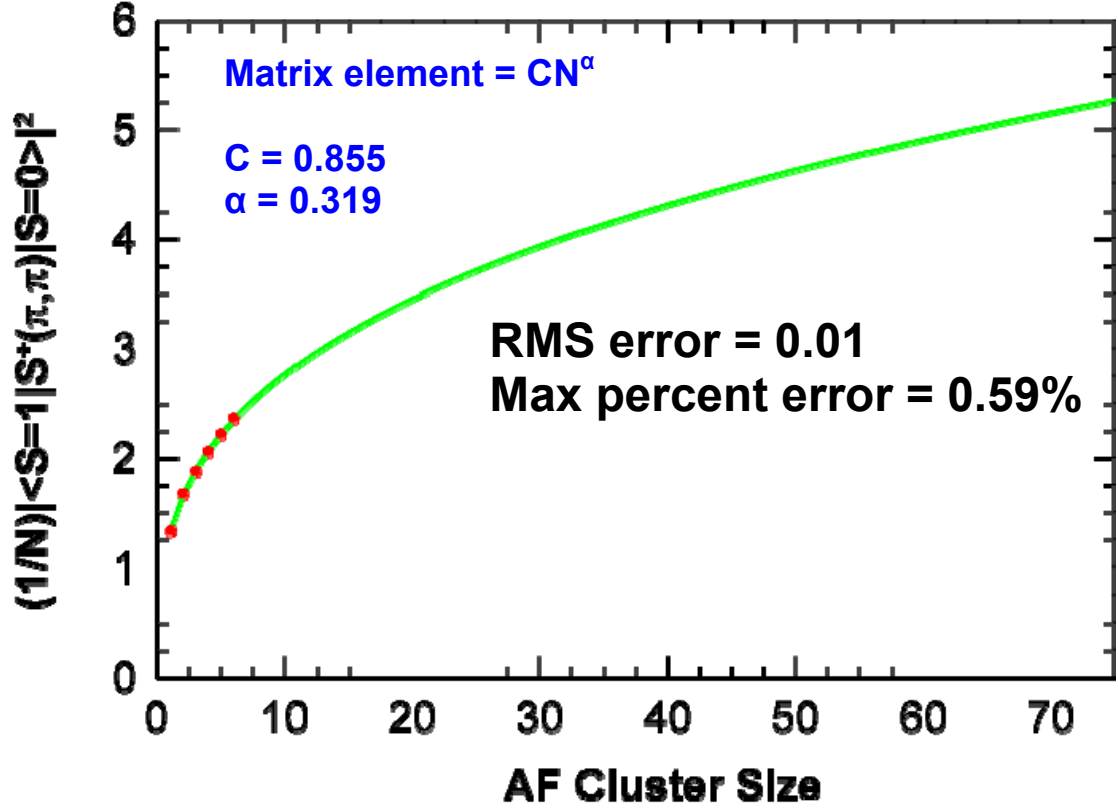

**Figure S5.** Calculated matrix element of the spin raising operator at momentum $(\pi, \pi)$ for all possible clusters up to 24 undoped AF $d^9$ spins. The curve is very well fit by the power law expression, $CN^\alpha$, where $C = 0.855$ and $\alpha = 0.319$. The RMS error is 0.01 and the maximum percentage error is 0.59%.

Finally, to compute the integrated susceptibility, we need to know how many clusters there are of each size in the material. We calculated the number of clusters of size n, $N_{clus}(n)$, by randomly doping 1000 ensembles of a 2000 X 2000 lattice. The results are shown in figure S6. The values were determined by averaging over 1000 ensembles. The figure shows the results for dopings $x = 0.16$ and $x = 0.10$. Only the $x = 0.16$ data is used in the main text.

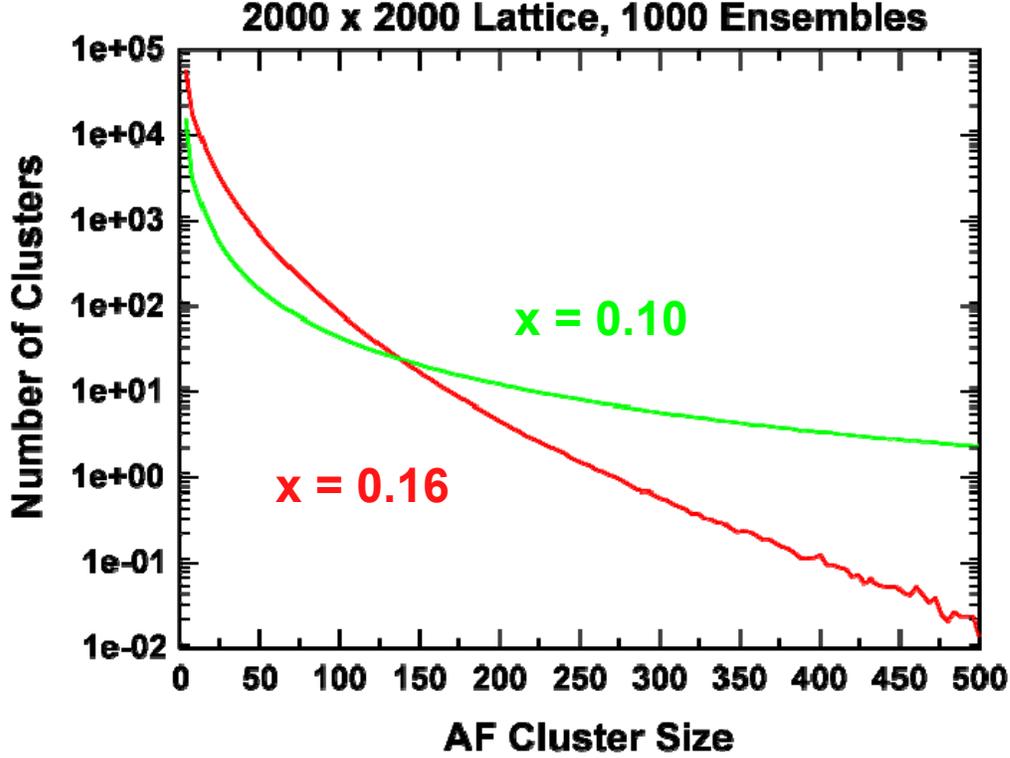

**Figure S6.** The number of clusters of each size for dopings of x = 0.16 and 0.10. For x = 0.16, there are fewer undoped $d^9$ sites leading to fewer large-sized clusters than x = 0.10. The curves in the figure were calculated for a 2000 x 2000 lattice and averaging was done over 1000 ensembles. The y-axis (number of clusters) is plotted on a log scale and is not normalized by the number of lattice sites. If the lattice size was doubled, the number of clusters would double.

**Calculation of neutron resonance width**

The S=1 state excited by the neutron decays by electron-spin scattering with the $x^2$-$y^2$/p$\sigma$ band electrons. The width of the resonance peak, $\Gamma$ (half-width at half-maximum, HWHM), is given by $\Gamma=\pi|W|^2 F_{BCS}$ where $F_{BCS}$ is the standard electron spin-flip scattering term with BCS coherence factors[2]. $|W|^2=|M|^2(1-\lambda)/2(1-\lambda)^{1/2}$ where $\lambda=[1-(E_{res}/2J_{dd})^2]^{1/2}$ and the matrix element $M \approx J_{dd}^3$. For the bilayer systems, $F_{BCS}$ only includes scattering between the bonding and anti-bonding bands.